\documentclass[sn-standardnature]{sn-jnl}

\jyear{2023}%

\theoremstyle{thmstyleone}%
\theoremstyle{thmstyletwo}%

\theoremstyle{thmstylethree}%

\newcommand{\AAM}{\text{\normalfont\AA}}

\usepackage{comment}

\newenvironment{newtxt}%
{\color{black}}%
{}

\begin{document}

\title[On the Cosmological Time Dilation of Quasars]{Detection of the Cosmological Time Dilation of 
High Redshift Quasars}

\author*[1]{\fnm{Geraint F.} \sur{Lewis}}\email{geraint.lewis@sydney.edu.au}

\author[2]{\fnm{Brendon J.} \sur{Brewer}}\email{bj.brewer@auckland.ac.nz}

\affil*[1]{\orgdiv{Sydney Institute for Astronomy, School of Physics, A28}, \city{The University of Sydney}, \state{NSW} \postcode{2006}, \country{Australia}}

\affil[2]{\orgdiv{Department of Statistics}, \orgname{The University of Auckland}, \orgaddress{\street{Private Bag 92019}, \city{Auckland} \postcode{1142}, \country{New Zealand}}}

\abstract{
A fundamental prediction of relativistic cosmologies is that, due to the expansion of space, observations of the distant cosmos should be time dilated and appear to run slower than events in the local universe. Whilst observations of cosmological supernovae unambiguously display the expected redshift-dependent time dilation, this has not been the case for other distant sources. Here we present the identification of cosmic time dilation in a sample of 190 quasars monitored for over two decades in multiple wavebands by assessing various hypotheses through Bayesian analysis. This detection counters previous claims that observed quasar variability lacked the expected redshift-dependent time dilation. Hence, as well as demonstrating the claim that the lack of the redshift dependence of quasar variability represents a significant challenge to the standard cosmological model, this analysis further indicates that the properties of quasars are consistent with them being truly cosmologically distant sources.
}

\keywords{}

\maketitle


\section{Introduction}\label{introduction}
A fundamental consequence of the relativistic picture of expanding space is cosmological time dilation, where 
events in the distant universe appear to run slowly compared to those in the local cosmos
\cite{1927ASSB...47...49L,1931MNRAS..91..483L,1939ApJ....90..634W}. 
Whilst this time dilation has been unambiguously detected in the light curves exhibited by cosmologically distant supernovae \cite{1996ApJ...466L..21L,1997AJ....114..722R,2001ApJ...558..359G,2005ApJ...626L..11F,2008ApJ...682..724B}, 
the appearance of time dilation in other cosmic sources is less conclusive. 
\begin{newtxt}
For example, whilst examinations of the light curves of gamma-ray bursts (GRBs) have generally shown consistency with the expected cosmological signature, uncertainties in the detailed emission mechanism and expected light curve characteristics mean that this detection has not been definitive
\cite[e.g.][]{1994ApJ...424..540N,2013ApJ...765..116K,2013ApJ...778L..11Z,2014MNRAS.444.3948L,2022JCAP...02..010S}. 
Furthermore, the role of the more recently discovered fast radio bursts \cite[FRBs:][]{2022A&ARv..30....2P} as 'standard clocks' is similarly limited 
by knowledge of the physical processes driving the output \cite[][]{2022arXiv221203972Z}.

Quasars have been known to be variable sources since their discovery in the 1960s \cite{1963Natur.197.1040S}, with emission arising from a relativistic accretion disk orbiting a supermassive black hole \cite{1964ApJ...140..796S}. 
However, it has been claimed
\end{newtxt} 
that the variability displayed by quasars over a broad range of redshifts does not show the expected cosmological time dilation 
\cite{1993Natur.366..242H,2001ApJ...553L..97H,2010MNRAS.405.1940H}. 
This has led to the suggestion that quasar variability is not intrinsic, but is due to microlensing due to the presence of cosmologically distributed black holes \cite{1997ApJ...482L...5H,2022MNRAS.512.5706H}. 
Others have stated that this points to more fundamental issues with our cosmological ideas 
\cite[e.g.][]{universe1030307,2017FoPh...47..711L,2017OAst...26..111C}, with even the suggestion that quasars are not cosmologically distant and that their observed redshifts are due to mechanisms other than the expansion of space.

In 2012, a study of the variability characteristics of  a sample of thirteen quasars observed behind the Magellanic Clouds as part of the MACHO microlensing program was suggestive of the expected $(1+z)$ time dilation dependence, where $z$ is the quasar redshift \cite{2012PhRvL.108w1302D}. However, with the small sample and relatively short monitoring period, this result is inconclusive. 
Recently, a new sample of the variability properties of quasars was presented  as 
part of the Dark Energy Survey 
and comprises 190 quasars, covering the redshift 
range $z\sim 0.2 \rightarrow 4.0$
\cite{2022MNRAS.514..164S}.
These are drawn from the sample of more than one hundred thousand spectroscopically
identified quasars with absolute magnitude $M\lt-22$ in the 300 $deg^2$ Sloan Digital Sky Survey of Stripe 82 (S82). 
Published as part of the SDSS DR7 quasar catalogue, the physical properties of these quasars are presented in \cite{Shen_2011}. This includes the bolometric 
luminosity which was determined through spectral fitting and correction from 
composite spectral energy distributions \cite{Richards_2006}. These quasars
were photometrically observed 
between 1998 and 2020, and so for more than two decades, through the combination of multiple epochs of exposures from SDSS, PanSTARRS-1, and the Dark Energy Survey, with additional follow-up monitoring with Blanco 4m/DECam. 

The total dataset consists of roughly two hundred photometric observations of each quasar in multiple bands, although the cadence of these observations is very uneven over the observing period. To account for this when calculating characteristic 
timescales of the quasar variabilities, \cite{2022MNRAS.514..164S}
adopted a Gaussian process regression \cite[e.g.][]{https://doi.org/10.48550/arxiv.2209.08940} to interpolate the photometric data and the associated uncertainties between the observations; details are given 
in Appendix {\bf A2} of their paper.
Each quasar light curve in each band is represented as a Damped Random Walk 
\cite[DRW;][]{2009ApJ...698..895K,10.1093/mnras/stac803}; this is found to be an accurate description of quasar variability with only a mild dependence on the 
physical properties of the quasars \cite{2021ApJ...907...96S}.
Practically, this defines the covariance matrix of the Gaussian Process that describes the variability. 
With this, the Gaussian Process regression software, {\tt Celerite} \cite{celerite}, is used to determine the characteristic DRW time scale, as well
as the $16^{th}$ and $84^{th}$ percentiles of the distribution. 
\begin{newtxt}
Armed with these bolometric luminosities and variability time scales 
drawn from the DRW analysis, the goal of this paper is to search for the signature of cosmological time dilation of these distant sources.
\end{newtxt}

\section{Results}\label{results}
In the following analysis, we consider the redshift dependence of time dilation to be 
of the form $(1+z)^n$, where $z$ is the redshift of the source. 
Clearly, for the expected cosmological dependence, $n=1$, whilst $n=0$ demonstrates no 
redshift dependence, representative of the claims from several previous studies of quasar samples \cite[e.g][]{2010MNRAS.405.1940H}. 
To explore the various possibilities, several distinct hypotheses are explored.
These are:
\begin{itemize}
\item ${\cal H}_0$: $n$ is fixed at zero, representing  no redshift dependence on the observed quasar timescales.
\item ${\cal H}_1$: $n$ is fixed at one, representing the expected redshift dependence of the cosmological time dilation.
\item ${\cal H}_2$: $n$ is treated as a free parameter.
\item ${\cal H}_3$ and ${\cal H}_4$: $n$ is fixed at $-1$ and $2$ respectively.
\end{itemize}
The final two cases represent extreme cases where additional influences, such as quasar evolution, may significantly 
influence the observed time variability of quasars.

As outlined in the Methodology (Section~\ref{methods}), these differing hypotheses were compared through the calculation
of the Bayesian evidence \cite{10.2307/2291091} for each situation under consideration, with the results of these calculations presented in Table~\ref{tab1}; 
in assessing the ratio of Bayesian evidences, a factor of 10-100 is considered a strong favouring of one hypothesis over another, 
whereas greater than 100 is decisive
\cite{doi:10.1080/01621459.1995.10476572}.
One immediate conclusion is that the favoured hypothesis is ${\cal H}_1$, the case where $n=1$, which represents the expected redshift dependence of the cosmological time dilation. 
This is significantly favoured over the
alternative ${\cal H}_0$, with an evidence ratio greater than $10^5$, 
which represents the situation where there is no redshift dependence on the observed 
timescales of cosmological variability. Furthermore, ${\cal H}_1$ is significantly favoured over the two 
extreme cases, ${\cal H}_3$ and ${\cal H}_4$.

\begin{figure}[h]%
\centering
\includegraphics[width=0.9\textwidth]{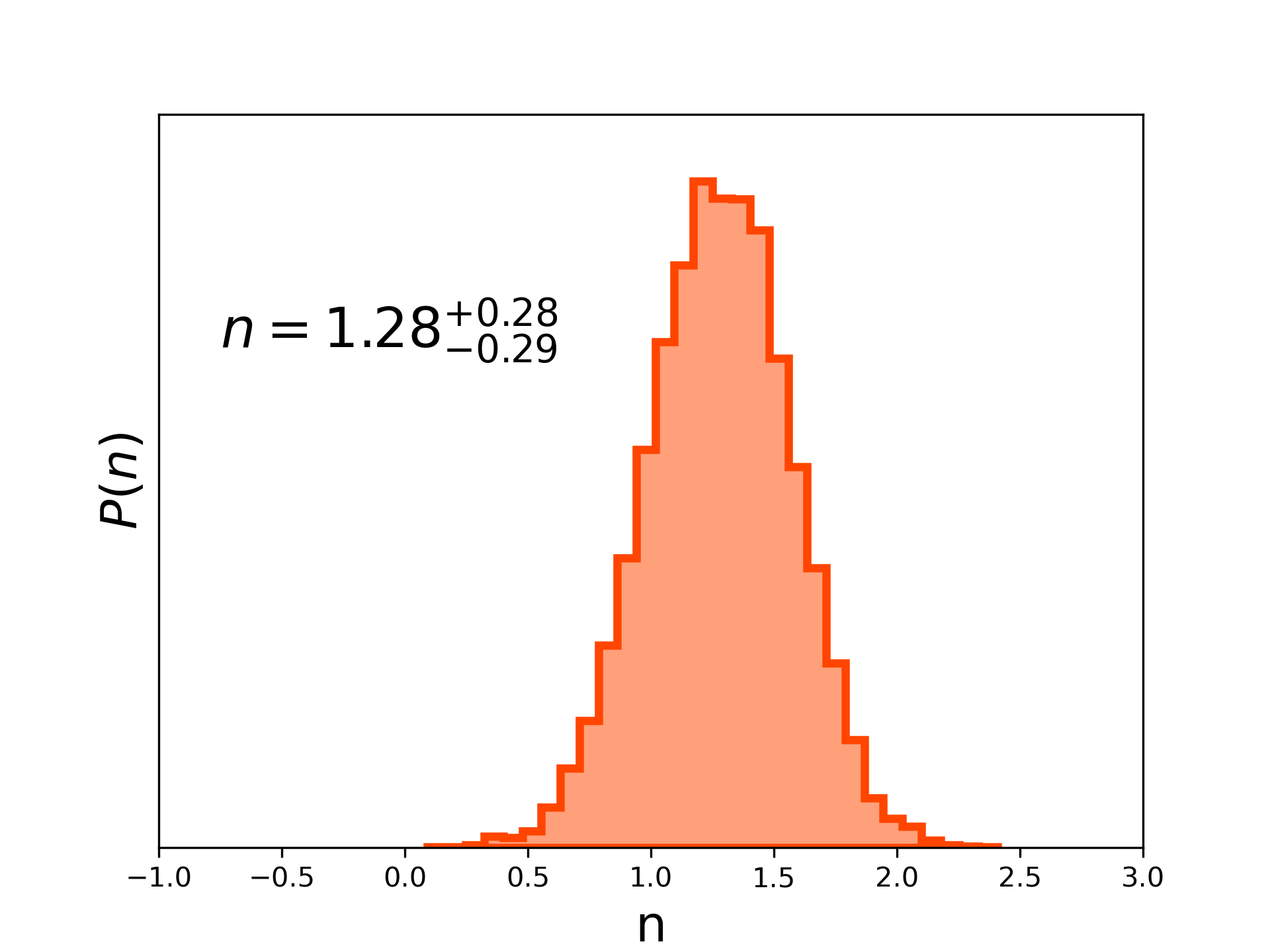}
\caption{The posterior distribution of $n$, where the redshift dependence of the observed time dilation
is given by $(1+z)^n$, for the Bayesian exploration of ${\cal H}_2$, where this index is treated as a free parameter in the analysis. From this distribution, $n = 1.28^{+0.28}_{-0.29}$, where the best fit value is taken as the
median ($50^{th}$ percentile), whilst the uncertainties represent the
$16^{th}$ and $84^{th}$ percentiles. This distribution was determined through an exploration of the posterior probability
space with {\tt DNest4} \cite{JSSv086i07}.
}\label{fig1}
\end{figure}

The posterior distribution for the redshift dependence for the time dilation, $n$, specifically ${\cal H}_2$, 
where this is treated as a free parameter, is presented in Figure~\ref{fig1}.
Reflecting the previous analysis, the is clearly offset from zero, indicative of a redshift dependence of the observed timescale of variability over the quasar sample. 
This posterior distribution, which may be summarised using $n = 1.28^{+0.28}_{-0.29}$, is consistent with the expected cosmological 
dependence with $n=1$, and the presented analysis significantly favours
the presence of cosmological time dilation of the observed quasar variability.

\begin{newtxt}
\section{Methodology}\label{methods}
\end{newtxt}
Probing the fundamental nature of our universe often calls upon standard rulers or candles to allow us to determine the 
influence of expansion on observable quantities. In hunting for cosmological time dilation, a standard clock with a measurable timescale is required. However, the challenge with objects such as quasars, and other cosmological sources such as gamma-ray bursts, is the  complexity
of the physical processes driving their variability.
For quasars, where variability arises in the stochastic processes in the relativistic disk orbiting a supermassive black hole, the resultant 
luminosity fluctuations  could potentially be dependent on a multitude of physical properties, including the mass of the central black hole, 
the degree of accretion, and the wavelength of the observations. 

To address this, the sample of quasars under consideration here was split into a number of subsamples of objects with similar intrinsic properties in terms of their bolometric luminosity and the rest wavelength of observations. The observations under consideration were taken in 
the $(g,r,i)$ wavebands, and for the purpose of this study, 
the rest wavelength in each of the observed wavebands is defined to be
\begin{equation}
\lambda_{g} = \frac{4720 {\rm \AAM}}{1+z}, \ \ \ \lambda_{r} = \frac{6415 {\rm \AAM}}{1+z}, \ \ \ \lambda_{i} = \frac{7835 {\rm \AAM}}{1+z}
\label{rest}
\end{equation}
where $z$ is the redshift of the quasar under consideration and the numerical values are representative of the observed wavebands.

\begin{figure}[h]%
\centering
\includegraphics[width=0.9\textwidth]{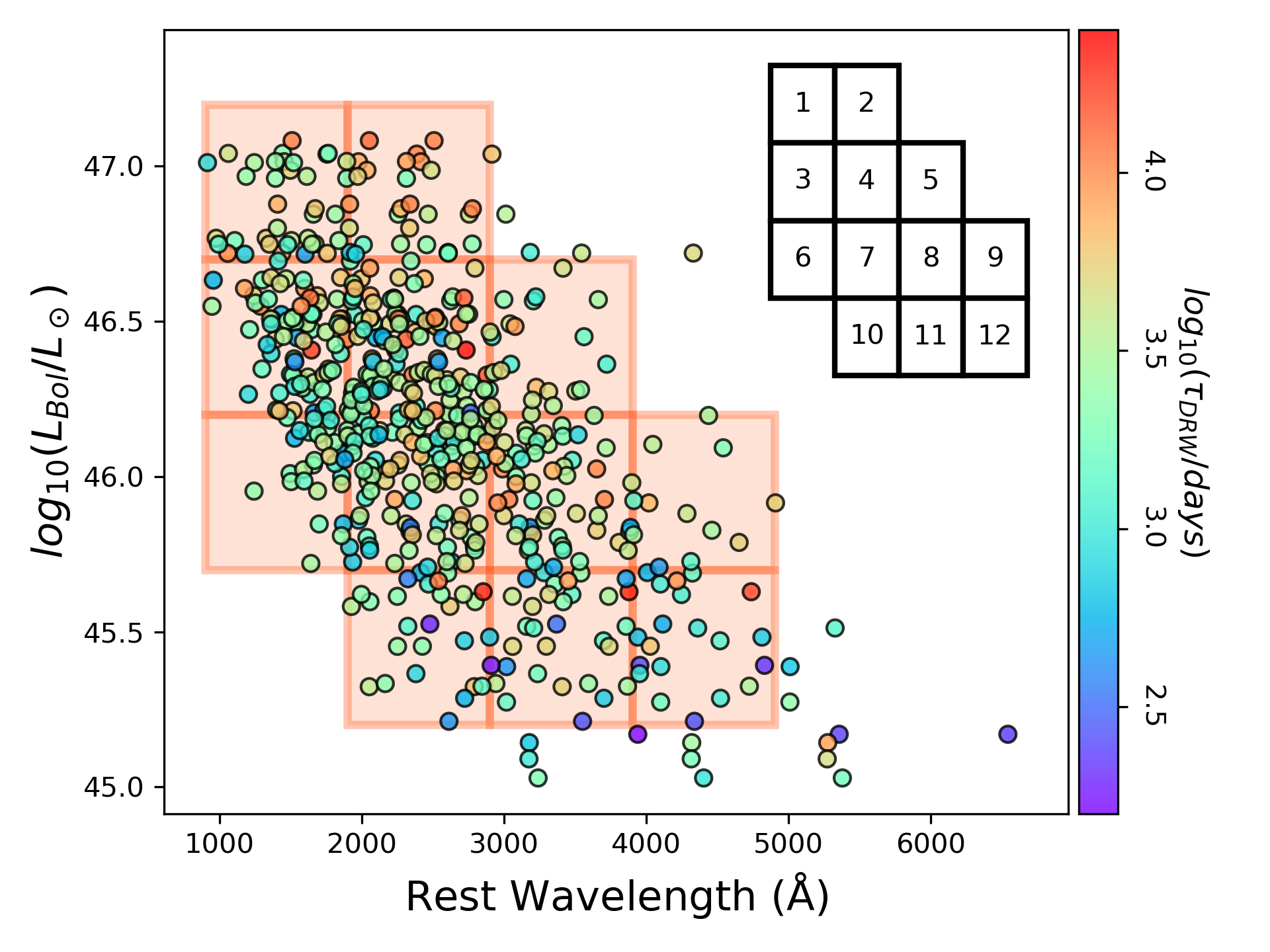}
\caption{The entire quasar sample under consideration as a function of rest wavelength and bolometric luminosity, colour-coded 
with the DRW timescale, $\tau_{DRW}$. The underlying rectangles in salmon pink represent the boundaries of the subsamples
employed in the analysis presented in this paper. Inset is the labelled numbers of the fields.}\label{fig2}
\end{figure}

The quasar subsamples are presented graphically in Figure~\ref{fig2}, which presents the rest wavelengths of each quasar in the $(g,r,i)$ bands versus their bolometric luminosity. Each quasar has been colour-coded with its variability timescale, $\tau_{DRW}$, assessed by fitting each observed light curve in each band with a damped random walk \cite[see][for more detail]{2022MNRAS.514..164S}. Underlying the quasar sample are the regions of twelve subsamples under consideration in the colour salmon. These were chosen to have a width in rest wavelength and bolometric luminosity of
$\Delta \lambda = 1000 {\rm \AAM}$ and
$\Delta \log({L_{Bol}/L_\odot}) = 0.5$.
Note that the regions are continuous, with the top-left subsample 
spanning $\Delta \lambda = 900 {\rm \AAM} \rightarrow 1900 {\rm \AAM}$, and 
$\Delta \log({L_{Bol}/L_\odot}) = 46.7 \rightarrow 47.2$; the details of the subsamples are given in Table~\ref{tab2}.
From Figure~\ref{fig2} it is clear that these subsamples encompass the majority of quasars presented in this survey,
and note that the combination of the three wavebands means that each subsample of quasars contains a
broader distribution of redshifts than if the wavebands were considered individually. Hence this combination provides 
a redshift lever arm which constrains the presence of cosmological time dilation in each subsample.
We also note that a 
 by-eye examination of Figure~\ref{fig2} is suggestive of a gradient in the variability timescale over the sample.

Given that in each subsample the quasars possess similar rest wavelengths and bolometric luminosities, we make the assumption that
they also possess the same characteristic intrinsic timescales, 
and hence any difference in timescale for a particular quasar subsample is due to the influence of cosmic time dilation and 
will show the appropriate dependence upon redshift.
Of course, 
the physics of quasar variability is likely to depend on a number of factors, and so this assumption is considering that these quasars will
exhibit similar variability properties in the mean;
we discuss this point again at the conclusion of this study.

For each quasar subsample (labelled with $k$), we model the observed variability 
timescales (i.e. $log_{10}( \tau_{DRW} / days)$) as 
\begin{equation}
M_k = C_k + n \log_{10} ( 1 + z )
\label{model}
\end{equation}
where $z$ is the redshift of the quasar and $n$ is the power of the expected cosmological term, that is $(1+z)^n$.
This model represents the variability with a different 
normalisation term, $C_k$, for each wavelength-luminosity bin, but demands the same cosmological dependence in terms of redshift. 

For the five distinct hypotheses considered, the normalisation terms, $C_k$, were allowed to vary, and so for the cases 
where $n$ is considered to be a fixed value, this corresponds to an exploration of a twelve-dimensional posterior probability
distribution. Physically, this situation reflects the situation where each wavelength-luminosity bin has a differing characteristic timescale, 
but a redshift dependence dependent upon the chosen value of $n$. 
For the remaining hypotheses, ${\cal H}_2$, where $C_k$ and $n$ are  treated as free parameters, this corresponds to a
thirteen-dimensional posterior probability distribution to be explored.

To calculate the Bayesian evidence (also known as marginal likelihood) for each of these hypotheses, it is
necessary to define a likelihood. It is important to note that the presented measurements and uncertainties of $\tau_{DRW}$ (in $\log_{10}$ space) are not 
symmetrical. Hence we represented the probability of each distribution
of $\log_{10}(\tau_{DRW}/days)$ as a skewed Gaussian, specifically 
{\tt scipy.stats.skewnorm} in the numerical approach which is written in python (represented as ${\cal SN}$).
The $16^{th}$, $50^{th}$ and $84^{th}$ percentiles of this skewed distribution were fitted to the given values via a straight-forward optimization. This resulted in uncertainties of these percentiles of typically less than $2-3\%$. Hence we can define the log of the likelihood as
\begin{equation}
\log {\cal L} = 
\sum_{k} \   
\sum_{l=g,r,i} \  
\sum_{m=1...N_q} \  
{\cal SN}.\textnormal{logPDF}({\cal M}_k , \theta_{l,m}( l,m \in k  ))
\label{likelihood}
\end{equation}
where $k$ sums over each of the subsample regions, $l$ over the observed wavebands and $m$ over the number of quasars, $N_q$, in the sample.
Also, $\theta_{l,m}$ are the parameters for the skewed Gaussian representing the probability distribution for $\log_{10}(\tau_{DRW}/days)$
for each of the quasars in each of the waveband, and $l,m \in k$ implies 
that the quasar should only be considered if its rest frame properties 
place it in subsample $k$.

\begin{figure}[ht]%
\centering
\includegraphics[width=0.9\textwidth]{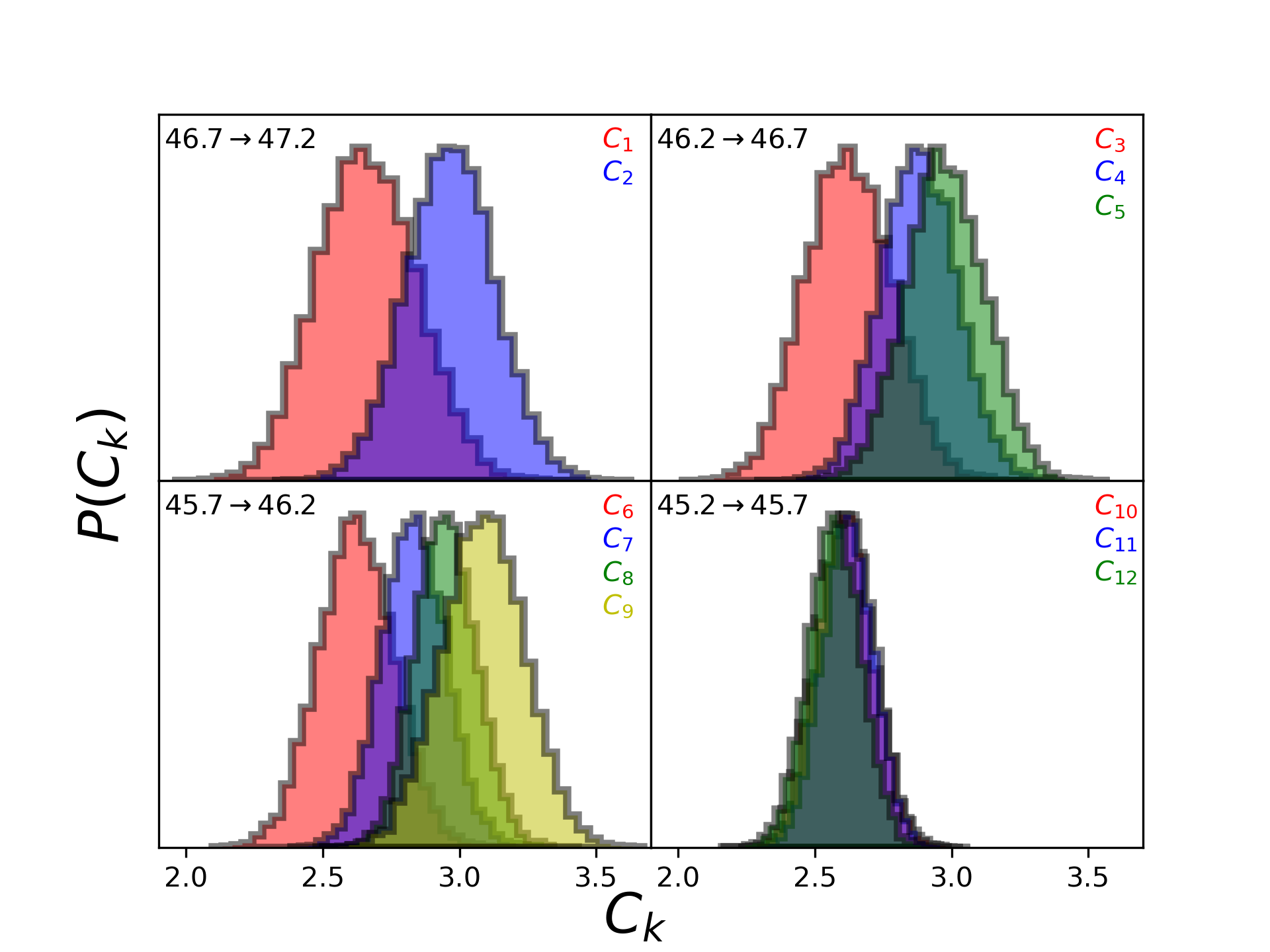}
\caption{The posterior distributions for the normalisation parameters, $C_k$, for  ${\cal H}_2$ where $n$ is treated as a free parameter. 
The bolometric luminosity range for each of the normalisation parameters is given in the upper left of each panel (see Table~\ref{tab2}).
As with Figure~\ref{fig1}, these were the result of the sampling of the posterior probability distribution with {\tt DNest4}.
}\label{fig3}
\end{figure}

To explore the posterior probability distribution and calculate the Bayesian evidence (also known as marginal likelihood), we employed Diffusive Nested Sampling {\tt DNest4} \cite{JSSv086i07}, a variant of the nested sampling technique \cite{2004AIPC..735..395S}. This allows for correct posterior sampling and marginal likelihood estimation even in the case where the constrained prior distributions are difficult to sample from or explore with Markov Chain Monte Carlo.
For simplicity, we use uniform priors over the normalisation parameters, $C_k$, between 1 and 5, and for ${\cal H}_2$ where $n$ is treated as a free parameter, a uniform distribution over $n$ is adopted between -1 and 3; the posterior distribution of $n$ for this hypothesis is presented in Figure~\ref{fig1}.
The normalisations, $C_k$, are well constrained and are  reproduced for completeness in Figure~\ref{fig3} for ${\cal H}_2$. 

\section{Conclusions}\label{conclusions}
This paper presents the detection of the cosmological dependence of the time dilation in a recent
sample of almost two hundred quasars. These were monitored in multiple wavebands over a two-decade period, allowing
the determination of a characteristic timescale by treating the observed quasar variability as a damped 
random walk. 

Through an assessment of the Bayesian evidence, it was found that the hypothesis considering the expected $(1+z)$ cosmological 
dependence provides a significantly better description of the data than the case where there is no dependence on redshift. 
In considering the redshift dependence of quasar variability to be of the form $(1+z)^n$, where $n$ is treated as a free 
parameter, the posterior distribution is found to be $n = 1.28^{+0.28}_{-0.29}$, again consistent with the expected 
cosmological
\begin{newtxt}
expansion of space. 
This detection of the cosmic expansion directly imprinted onto the variability of quasars further demonstrates that their observed properties are consistent with them being luminous and variable sources at cosmological distances, and counters previous claims that  quasar variability is not intrinsic, but instead is due to external influences or non-standard physics. 
This has an immediate impact on various claims, such as the presence of a cosmologically significant population of microlensing black holes \cite[e.g.][]{1993Natur.366..242H,2022MNRAS.512.5706H} or more esoteric ideas about the framework of the universe \cite[][]{2022IJMPD..3150084S}, and is further evidence that we inhabit an expanding relativistic universe.
\end{newtxt}

\begin{newtxt}
We do note that our result of $n = 1.28^{+0.28}_{-0.29}$ could be consistent with an offset from the expected cosmological value of $n=1$ and could potentially indicate the presence of additional factors such as 
an evolution of quasars over cosmic time in addition to the time dilation due to cosmic expansion.
Of course, we could imagine that quasar evolution over cosmic time could be responsible for the observed redshift dependence of the DRW time scale, but as we are considering similar quasars in terms of the bolometric luminosity 
and observed rest wavelength, it would be a curious coincidence for this evolution
to result in a $(1+z)$ dependence to spoof cosmic expansion. Furthermore, if quasar evolution were solely responsible for the observed DRW properties then the resulting lack of the expected cosmic time dilation would present a severe challenge to our cosmological model. However, it is important to note that there are some potential correlations of the DRW timescales with the inferred intrinsic properties of the quasars \cite[e.g.][]{2021ApJ...907...96S}, although these are not strong, and more extensive photometric datasets in terms of the number of quasars and the duration of their photometric lightcurves will be required to cleanly separate the influence of cosmic expansion from quasar evolution.
\end{newtxt}

In closing, we note that the lack of detection of the time dilation of quasar variability in previous studies 
\begin{newtxt}
is potentially due to the relatively small sample size in terms of the number of quasars under consideration \cite{2012PhRvL.108w1302D}, or the cadence of data sampling and characterisation of the quasar variability \cite{2001ApJ...553L..97H,2010MNRAS.405.1940H}.
Built on the observations of Stone et al. (2022) \cite{2022MNRAS.514..164S}, this present study has demonstrated that we are now in an epoch where we have observations of a sufficiently large number of quasars spanning a broad range in redshifts, and observed over extended periods and with a cadence that overcomes their stochastic nature and  results in an accurate characterisation of their variability, yielding a robust determination of the imprint of cosmological expansion on their light curves.
\end{newtxt}
Furthermore, with upcoming programs such as the Vera Rubin Observatory Legacy Survey of Space and Time (LSST), the number of quasars observed at high temporal cadence will rapidly increase and the measurement of cosmological time 
dilation, and potentially the influence of quasar evolution, will become readily observable \cite[e.g.][]{10.1093/mnras/stac803}.

\backmatter

\bmhead{Data Availability}
The source data for this project is available at {\tt https://zenodo.org/record/ 5842449\#.YipOg-jMJPY}, with the details 
of the available FITS tables presented in Stone et al. (2022) \cite{2022MNRAS.514..164S}. Note that a revised version of this catalogue was recently released due to an error in some rest frame quantities. This revision does not impact any of the research presented in this paper.
The software for this project is available at
{\tt https://github.com/eggplantbren/QuasarTimeDilation}.

\bmhead{Code Availability}
This project made use of several publicly available software packages, especially 
{\tt DNest4} \cite{JSSv086i07} to undertake the exploration of the posterior probability space and 
calculate the Bayesian evidence by integrating across this space. 
Further software packages employed include
{\tt matplotlib} \cite{Hunter:2007}, {\tt numpy} \cite{harris2020array}, {\tt scipy} \cite{2020SciPy-NMeth}. Initial explorations of the posterior probability space were undertaken with {\tt emcee} with corner plots prepared with {\tt corner} \cite{corner}.
The software employed as part of this project will be made available on reasonable request to the corresponding authors.

\bmhead{Acknowledgements}
We thank Stone et al. (2022) \cite{2022MNRAS.514..164S} for making their data and the results of their analysis publicly available. We also thank Scott Croom for his input and advice on quasar variability surveys.
We further thank the teams responsible for creating and maintaining the various software packages, detailed below, that this
study has employed.
GFL would like to thank the hospitality of the Lowell Observatory where the last stages of this work were completed during a period of isolation due to the contraction of covid.

\begin{newtxt}
\bmhead{Author Contribution Statement}
The project was  conceived by GFL, including an initial exploration of the data, the definition of the models and hypotheses considered, the likelihood function and sampling of the posterior space. BJB undertook detailed sampling and calculating the Bayesian evidence using {\tt DNest4}. Both authors discussed the results of the exploration in detail and determined the resulting conclusion. Both were responsible for the writing of the manuscript. 
\end{newtxt}

\begin{newtxt}
\bmhead{Competing Interests Statement}
The authors declare no competing interests.
\end{newtxt}

\newpage

\bmhead{Tables}
$ $

\begin{table}[h]
\begin{center}
\begin{minipage}{174pt}
\caption{The marginal likelihoods for the various hypotheses considered in this paper. 
As described in more detail in the Methods (Section~\ref{methods}), these were calculated with the
diffusive nested sampling approach {\tt DNest4} \cite{JSSv086i07}.}\label{tab1}%
\begin{tabular}{@{}crcr@{}}
\toprule
Hypothesis & $n$ & $\log{\cal Z}$  & ${\cal Z} / {\cal Z}_{max}$ \\
\midrule
${\cal H}_0$  & 0  &  -366.12 &  $9.3\times 10^{-6}$\\
${\cal H}_1$  & 1  & -354.53  &  1  \\
${\cal H}_2$  & Free  & -356.52  &  0.14 \\
${\cal H}_3$  & -1  & -390.13  &  $3.5 \times 10^{-16}$  \\
${\cal H}_4$  & 2  &  -358.36 & $2.2 \times 10^{-2}$  \\
\botrule
\end{tabular}
\end{minipage}
\end{center}
\end{table}

\newpage

\begin{table}[h]
\begin{center}
\begin{minipage}{340pt}
\caption{The properties of the survey subsamples presented in Figure~\ref{fig2}, with the boundaries of the
subsamples given by $\Delta\lambda$, rest wavelength, and $\Delta \log_{10}({L_{Bol}/L_\odot})$, bolometric luminosity. 
The remaining columns give the number 
of quasar light curves in each subsample, $N_{qs}$, as well as the redshift range of those quasars, $\Delta z$, 
and timescale for the observed variability as given by treating this as a damped random walk, 
$\Delta \log_{10}(\tau_{DRW}/days)$.
}\label{tab2}%
\begin{tabular}{@{}cccccc@{}}
\toprule
Subsample & $\Delta\lambda$ (\AA) &  $\Delta \log({L_{Bol}/L_\odot})$ & $N_{qs}$ & $\Delta z$ & $\Delta \log_{10}(\tau_{DRW}/days)$\\
\midrule
1  &$\ 900\rightarrow 1900$&$46.7\rightarrow47.2$& 37 & $1.60\rightarrow4.15$ & $2.68\rightarrow4.03$ \\
2  & $1900\rightarrow 2900$&$46.7\rightarrow47.2$& 27 & $0.81\rightarrow3.00$ & $2.83\rightarrow4.11$ \\
3  &$\ 900\rightarrow 1900$&$46.2\rightarrow46.7$& 74 & $1.55\rightarrow3.98$ & $2.61\rightarrow4.23$ \\
4  & $1900\rightarrow 2900$&$46.2\rightarrow46.7$&111 & $1.11\rightarrow3.01$ & $2.49\rightarrow4.42$ \\
5  & $2900\rightarrow 3900$&$46.2\rightarrow46.7$& 22 & $1.11\rightarrow1.70$ & $3.03\rightarrow3.93$ \\
6  &$\ 900\rightarrow 1900$&$45.7\rightarrow46.2$& 30 & $1.48\rightarrow2.80$ & $2.70\rightarrow3.71$ \\
7  & $1900\rightarrow 2900$&$45.7\rightarrow46.2$&101 & $0.68\rightarrow2.80$ & $2.55\rightarrow3.92$ \\
8  & $2900\rightarrow 3900$&$45.7\rightarrow46.2$& 58 & $0.60\rightarrow1.69$ & $2.63\rightarrow4.03$ \\
9  & $3900\rightarrow 4900$&$45.7\rightarrow46.2$& 11 & $0.60\rightarrow1.00$ & $2.66\rightarrow3.84$ \\
10 & $1900\rightarrow 2900$&$45.2\rightarrow46.7$& 27 & $0.63\rightarrow1.45$ & $2.30\rightarrow4.30$ \\
11 & $2900\rightarrow 3900$&$45.2\rightarrow46.7$& 31 & $0.47\rightarrow1.45$ & $2.27\rightarrow4.30$ \\
12 & $3900\rightarrow 4900$&$45.2\rightarrow46.7$& 20 & $0.47\rightarrow0.98$ & $2.33\rightarrow4.20$ \\
\botrule
\end{tabular}
\end{minipage}
\end{center}
\end{table}

\newpage

\bmhead{Figures}
$ $

\newpage

\newpage

\newpage

\bibliography{paper}


\end{document}